\documentclass[preprint,pra,showpacs,nofootinbib]{revtex4}
\usepackage[mathscr]{euscript}
\usepackage[dvips]{pstcol}
\usepackage{graphicx}
\usepackage{float,epsfig}
\usepackage{dcolumn}
\usepackage{bm}
\usepackage{graphicx}
\usepackage{dcolumn}
\usepackage{amsmath,amssymb,amsthm}
\usepackage{subfigure}
\usepackage[colorlinks=true,linkcolor=blue]{hyperref}
\newcommand{\bea}{\begin{eqnarray}}
\newcommand{\eea}{\end{eqnarray}}
\newcommand{\beq}{\begin{equation}}
\newcommand{\eeq}{\end{equation}}

\def\/{\over}

\begin{document}

\title{Protection of entanglement between two two-level atoms }
\author{Anwei Zhang$^{1,2}$}

\affiliation{$^{1}$Institute of Physics and Key Laboratory of Low
Dimensional Quantum Structures and Quantum
Control of Ministry of Education,\\
Hunan Normal University, Changsha, Hunan 410081, China \\
$^{2}$Department of Physics and Astronomy, Shanghai Jiao Tong University, Shanghai 200240, China}


\begin{abstract}
The dynamical evolution of entanglement between two
polarizable two-level atoms in weak interaction with electromagnetic vacuum fluctuations is investigated. We find that, for initial Bell state $\psi^{\pm}$,
the decay rate of entanglement between atoms is just the superradiant spontaneous emission rate, which
depends not only on the spontaneous emission rate of atom but also on the
modulation of the spontaneous emission rate due to the presence of another atom. It is shown that, with the presence of a boundary, the entanglement between transversely polarizable atoms can  be protected for a very long time. It is pointed out that when the two atoms in initial Bell state $\psi^{-}$ are put close enough , the entanglement between them can also be protected.
\end{abstract}
\pacs{03.67.Pp, 03.67.Mn, 03.65.Yz}

\maketitle

Entanglement is one of the most interesting
and prominent phenomena which distinguish the classical and quantum worlds. It has been recognized as
a resource for quantum communication and teleportation, as well as computational tasks \cite{bou}.
In practice, a main obstacle for entanglement in the realization of quantum technologies is the inevitable interactions between system and environment which can lead to decoherence. Therefore, how entanglement between atoms in external environment evolve and how to avoid the influence of environment on entanglement become important issues in quantum information science.

In the past few years, many proposals have been suggested for fighting against the deterioration of entanglement under the impact of environment, such as,
decoherence-free subspaces \cite{dec1,dec2,dec3}, quantum error correction code \cite{qec1,qec2,qec3}, dynamical decoupling \cite{dd1,dd2,dd3} and quantum
Zeno dynamics \cite{zeno,zeno1,zeno2}. However, when the time scale characterizing the undesired interaction is too short, dynamical decoupling will not work
due to the lack of memory \cite{13,14}. And the efficiency of Zeno dynamics is restricted by the requirement of high measurement frequency.

Recently it was found that, with the presence of a boundary, the quantum Fisher information of the parameters of the initial atomic state
can be shielded from the influence of the vacuum fluctuations in certain circumstances as if it were a closed system \cite{yao}. And it was shown that
quantum coherence of a two-level atom in the presence of a boundary could be effectively inhibited when the atom is transversely polarizable and near the boundary \cite{ta,ta2}.

It is nature for us to wonder whether entanglement between two atoms can be protected for a long time in the presence of a reflecting plate.
By investigating the dynamical evolution of entanglement between two two-level atoms interacting with electromagnetic vacuum fluctuations, we demonstrate that, with the presence of a boundary, the entanglement can be indeed protected from decreasing as if it were isolated from environment. Since two atoms are considered and the existence of one atom will have an influence on another atom, so we also wonder how this influence is reflected in the
 evolution of entanglement. Besides we will also investigate the role of atomic polarizations in protecting of entanglement.

Let us consider two identical two-level atoms interacting with fluctuating electromagnetic fields in vacuum, in such a case, the total Hamiltonian
of the coupled system can be described by $H=H_s+H_f+H_I$. Here $H_s$ is the free
Hamiltonian of the two atoms and its explicit expression is given by
$H_s=\sum^{2}_{j=1}\hbar\omega_0\sigma^{+}_j\sigma^{-}_j,$
where $\omega_0$ is the level spacing of atom, $\sigma^{+}_j=|e_j\rangle\langle g_j|$ and $\sigma^{-}_j=|g_j\rangle\langle e_j|$ are, respectively, the raising and lowering operators of the atom $j$.
$H_f$ is the free Hamiltonian
of the quantum field which takes the form
$H_f=\sum_{k\lambda}\hbar\omega_{\vec{k}} a^{\dag}_{\vec{k}\lambda}a_{\vec{k}\lambda}.$
Here $a^{\dag}_{\vec{k}\lambda}$, $a_{\vec{k}\lambda}$ are the creation and annihilation operators for a photon with momentum $\vec{k}$, frequency $\omega_{\vec{k}}$ and polarization $\lambda$.
Finally, the atom-field interaction Hamiltonian $H_I$ can be written in the electric dipole approximation
\begin{equation}\label{3}
 H_I=-e\sum^{2}_{j=1}\vec{r}_j\cdot\vec{E}(\vec{x}_j)=-e\sum^{2}_{j=1}
 (\vec{d}_j\sigma^{+}_j+\vec{d}^{\ast}_j\sigma^{-}_j)\cdot
 \vec{E}(\vec{x}_j),
\end{equation}
where $e$ is the electron electric charge, $e\vec{r}_j$ is the electric dipole moment for atom $j$, $\vec{d}_j=\langle e_j|\vec{r}_j|g_j\rangle$, and $\vec{E}(\vec{x}_j)$ is the electric field strength evaluated at the position $\vec{x}_j$ of atom $j$.

The time evolution of the system is governed by the Schr\"{o}dinger equation which in the interaction picture has the form
\begin{equation}\label{6}
 i\hbar\partial_t|\varphi(t)\rangle=H_I(t)|\varphi(t)\rangle,
\end{equation}
where $H_I(t)=-e\sum^{2}_{j=1}(\vec{d}\sigma^{+}_je^{i\omega_0t}
+\vec{d}^{\ast}\sigma^{-}_je^{-i\omega_0t})\cdot
 \vec{E}(\vec{x}_j, t)$, $\vec{E}(\vec{x}_j, t)=e^{iH_ft/\hbar}\vec{E}(\vec{x}_j)e^{-iH_ft/\hbar}$ and we have let $\vec{d}_1=\vec{d}_2=\vec{d}$ for simplicity.
 Now decomposing $\vec{E}(\vec{x}_j,t)$ in $H_I(t)$ into positive- and negative-frequency parts:
$ \vec{E}(\vec{x}_j,t)=\vec{E}^{+}(\vec{x}_j,t)+\vec{E}^{-}(\vec{x}_j,t)$
with $\vec{E}^{+}(\vec{x}_j, t)|0\rangle=0$ and $\langle0|\vec{E}^{-}(\vec{x}_j, t)=0$, we have $H_I(t)$ in rotating-wave approximation
\begin{equation}\label{rot}
H_I(t)=-e\sum^{2}_{j=1}\big(\vec{d}\cdot\vec{E}^{+}(\vec{x}_j, t)\sigma^{+}_je^{i\omega_0t}
+\vec{d}^{\ast}\cdot
 \vec{E}^{-}(\vec{x}_j, t)\sigma^{-}_je^{-i\omega_0t}\big).
\end{equation}

Taking the initial states of atoms as the Bell state $\psi^{+}$ and environment as vacuum state respectively,
  \begin{equation}\label{7}
 |\varphi(0)\rangle=\frac{1}{\sqrt{2}}(|e_1g_2\rangle+|g_1e_2\rangle)|0\rangle.
 \end{equation}
 The state vector at time $t$ can be written as
 \begin{equation}\label{8}
|\varphi(t)\rangle=b_1(t)|e_1g_2\rangle|0\rangle+b_2(t)|g_1e_2\rangle|0\rangle
+\sum_{k\lambda}b_{\vec{k}\lambda}(t)|g_1g_2\rangle|1_{\vec{k}\lambda}\rangle,
 \end{equation}
with $|1_{\vec{k}\lambda}\rangle$ denoting one photon in the mode $(\vec{k},\lambda)$.
Now inserting (\ref{8}) and (\ref{rot}) into (\ref{6}), we can obtain
\begin{eqnarray}\label{9}
  \dot{b}_1(t)|0\rangle &=& \frac{ie}{\hbar}\vec{d}\cdot\vec{E}^{+}(\vec{x}_1,t)e^{i\omega_0t}
  \sum_{k\lambda}b_{\vec{k}\lambda}(t)|1_{\vec{k}\lambda}\rangle,\nonumber \\
  \dot{b}_2(t)|0\rangle &=& \frac{ie}{\hbar}\vec{d}\cdot\vec{E}^{+}(\vec{x}_2,t)e^{i\omega_0t}
  \sum_{k\lambda}b_{\vec{k}\lambda}(t)|1_{\vec{k}\lambda}\rangle,
  \end{eqnarray}
  and
  \begin{eqnarray}\label{10}
  \sum_{k\lambda}\dot{b}_{\vec{k}\lambda}(t)|1_{\vec{k}\lambda}\rangle &=& \frac{ie}{\hbar}b_1(t)e^{-i\omega_0t}\vec{d}^{\ast}\cdot\vec{E}^{-}(\vec{x}_1,t)|0\rangle
  + \frac{ie}{\hbar}b_2(t)e^{-i\omega_0t}\vec{d}^{\ast}\cdot\vec{E}^{-}(\vec{x}_2,t)|0\rangle.
  \end{eqnarray}
Integrating both side of (\ref{10}) over time, and substituting the result into (\ref{9}), we get
\begin{eqnarray}\label{11}
\dot{b}_1(t)&=&-\frac{e^{2}}{\hbar^{2}}\sum^{3}_{i,j=1}d_{i}d^{\ast}_{j}\int^{t}_0dt^{'}e^{i\omega_0(t-t^{'})}
 [\langle0|E^{+}_i(\vec{x}_1,t)E^{-}_j(\vec{x}_1,t^{'})|0\rangle b_1(t^{'})+\langle0|E^{+}_i(\vec{x}_1,t)E^{-}_j(\vec{x}_2,t^{'})|0\rangle
 b_2(t^{'})], \nonumber\\
 \dot{b}_2(t)&=& -\frac{e^{2}}{\hbar^{2}}\sum^{3}_{i,j=1}d_{i}d^{\ast}_{j}\int^{t}_0dt^{'}e^{i\omega_0(t-t^{'})}
 [\langle0|E^{+}_i(\vec{x}_2,t)E^{-}_j(\vec{x}_1,t^{'})|0\rangle b_1(t^{'})+\langle0|E^{+}_i(\vec{x}_2,t)E^{-}_j(\vec{x}_2,t^{'})|0\rangle
 b_2(t^{'})]. \nonumber\\
\end{eqnarray}
Then we can take Laplace transformation of (\ref{11}) to have
\begin{eqnarray}\label{12}
  s\tilde{b}_1(s)-\frac{1}{\sqrt{2}} &=& -L_{11}(s)\tilde{b}_1(s)
  -L_{12}(s)\tilde{b}_2(s), \nonumber\\
 s\tilde{b}_2(s)-\frac{1}{\sqrt{2}} &=& -L_{21}(s)\tilde{b}_1(s)
  -L_{22}(s)\tilde{b}_2(s),
\end{eqnarray}
where $\tilde{b}_{1/2}(s)=\int^{\infty}_0dtb_{1/2}(t)e^{-st}$, $L_{ab}(s)=\frac{e^{2}}{\hbar^{2}}\sum^{3}_{i,j=1}d_{i}d^{\ast}_{j}\int^{\infty}_0dte^{i\omega_0t-st}
 \langle0|E_i(\vec{x}_a,t)E_j(\vec{x}_b,0)|0\rangle$.

  Consider the two atoms separated by a distance $r$ and put at the same side of the boundary, which is located at $z=0$. The trajectories of atoms can be described by
$ \vec{x}_1 = (x_0, y_0, z_0),$
$\vec{x}_2 = (x_0+r, y_0, z_0).$
Then the electric field correlation function for the trajectories can be calculated to get \cite{jhep}
\begin{eqnarray}\label{13}
\langle0|E_x(\vec{x}_1, t)E_x(\vec{x}_1, 0)|0\rangle &=& \langle0|E_x(\vec{x}_2, t)E_x(\vec{x}_2, 0)|0\rangle\nonumber\\
&=& \frac{\hbar c}{\pi^{2}\varepsilon_0}\bigg[\frac{1}{(ct-i\varepsilon)^{4}}-\frac{c^{2}t^{2}+4z^{2}_0}{
[(ct-i\varepsilon)^{2}-4z^{2}_0]^{3}}\bigg],\nonumber\\
\langle0|E_x(\vec{x}_1, t)E_x(\vec{x}_2, 0)|0\rangle  &=& \langle0|E_x(\vec{x}_2, t)E_x(\vec{x}_1, 0)|0\rangle\nonumber\\
&=&\frac{\hbar c}{\pi^{2}\varepsilon_0}\bigg[\frac{1}{[(ct-i\varepsilon)^{2}-r^{2}]^{2}}
-\frac{c^{2}t^{2}+4z^{2}_0-r^{2}}
{[(ct-i\varepsilon)^{2}-r^{2}-4z^{2}_0]^{3}}\bigg].
\end{eqnarray}
 For simplicity, here we first consider the situation that the polarizations of the two atoms are all along the $x$-axis. In such a case, $L_{ab}(s)=\frac{e^{2}}{\hbar^{2}}d^{2}_{x}\int^{\infty}_0dte^{ i\omega_0t-st}
 \langle0|E_x(\vec{x}_a,t)E_x(\vec{x}_b,0)|0\rangle$. And it can be seen from (\ref{13}) that $L_{11}(s)=L_{22}(s)$, $L_{12}(s)=L_{21}(s)$. Thus we can obtain from (\ref{12})
\begin{eqnarray}
 \tilde{b}_1(s) =\tilde{b}_2(s)=\frac{1}{\sqrt{2}}\frac{1}{s+L_{11}(s)+ L_{12}(s)}.
 \end{eqnarray}
Then taking inverse Laplace transformation, one gets
 \begin{eqnarray}\label{b2}
 b_{1}(t) =b_{2}(t)=\frac{\sqrt{2}}{4\pi i}\int^{\infty}_{-\infty}\frac{e^{izt}dz}{z-iL_{11}(iz)- iL_{12}(iz)}.
 \end{eqnarray}

 We assume the interaction between the atoms and the field to be weak. So the Wigner-Weisskopf approximation can be adopted by neglecting the $z$ dependence of $L_{ab}(iz)$. Thus one has
 \begin{eqnarray}\label{c11}
L_{ab}(0)= \frac{e^{2}}{\hbar^{2}}d^{2}_x\big[\frac{1}{2}\mathcal{G}_{ab}(\omega_0)+
  i\mathcal{K}_{ab}(\omega_0)\big]
 \end{eqnarray}
with
\begin{eqnarray}\label{b1}
\mathcal{G}_{ab}(\omega_0)&=&\int^{\infty}_{-\infty}dte^{ i\omega_0 t}
 \langle0|E_x(\vec{x}_a,t)E_x(\vec{x}_b,0)|0\rangle,\nonumber\\
\mathcal{K}_{ab}(\omega_0)&=&-\frac{P}{2\pi }\int^{\infty}_{-\infty}\frac{\mathcal{G}_{ab}(\omega)}{\omega-\omega_0}d\omega.
\end{eqnarray}
Here $P$ denotes the Cauchy principal value. Note that $\frac{e^{2}}{\hbar^{2}}d^{2}_{x}\mathcal{G}_{11}(\omega_0)=\gamma_{11}$ is the spontaneous decay rate of two-level atom \cite{aw}, $\frac{e^{2}}{\hbar^{2}}d^{2}_{x}\mathcal{G}_{12}(\omega_0)=\gamma_{12}$
is the modulation of the spontaneous emission rate of one atom due to the presence of another atom \cite{dip},
$\frac{e^{2}}{\hbar^{2}}d^{2}_{x}\mathcal{K}_{11}(\omega_0)$ corresponds to level shift of the two-level atom
which can be neglected since it is irrelevant to our purposes and $\frac{e^{2}}{\hbar^{2}}d^{2}_{x}\mathcal{K}_{12}(\omega_0)=V$
is the dipole-dipole interaction potential.
Then from (\ref{b2}) and (\ref{c11}) , the state probability amplitudes can be obtained
\begin{eqnarray}\label{c1}
  b_1(t) =b_2(t)= \frac{\sqrt{2}}{2}e^{-\frac{1}{2}(\gamma_{11}+ \gamma_{12}+ 2iV)t}.
\end{eqnarray}

Now let us investigate the dynamics of entanglement between the two atoms.
We take concurrence \cite{woo} as a measure of entanglement, which is defined by
$\mathcal{C}=\mathrm{max}\{0, \sqrt{\lambda_1}-\sqrt{\lambda_2}-\sqrt{\lambda_3}-\sqrt{\lambda_4}\}$,
where $\lambda_i$ are the eigenvalues (in descending order) of the Hermitian matrix $\rho\tilde{\rho}$, in which $\tilde{\rho}=\sigma_y\otimes\sigma_y \rho^{\ast}\sigma_y\otimes\sigma_y$ and $\sigma_y$ is a pauli matrix. The concurrence is $1$ for
the maximally entangled states and $0$ for separable states.
The reduced density matrix $\rho$, which is obtained by tracing the density matrix of the total system over the field degrees of freedom, can be written in the basis of the product states, $|1\rangle=|e_1e_2\rangle$, $|2\rangle=|e_1g_2\rangle$, $|3\rangle=|g_1e_2\rangle$, $|4\rangle=|g_1g_2\rangle$. In this basis, the concurrence is
$\mathcal{C}(t)=2\mathrm{max}\{0, |\rho_{23}|\}=2|b_1(t)b^{\ast}_2(t)|$ \cite{ikram}.
Applying (\ref{c1}), we can find
\begin{equation}\label{22}
\mathcal{C}(t)=e^{-(\gamma_{11}+ \gamma_{12})t},
\end{equation}
where $\gamma_{ab}$ can be get by inserting (\ref{13}) into (\ref{b1})
\begin{eqnarray}
 \gamma_{11}&=&\gamma^{(0)}_{11}-3\gamma^{(0)}_{11}\frac{Z \cos Z+(Z^{2}-1)\sin Z}{2Z^{3}},\nonumber\\
\gamma_{12}&=&3\gamma^{(0)}_{11}\frac{\sin R
   -R\cos R}{R^{3}}-\frac{3\gamma^{(0)}_{11}}
  {2(R^{2}+Z^{2})^{3/2}}\bigg[\frac{Z^{2}-2R^{2}}{\sqrt{R^{2}+Z^{2}}} \cos\sqrt{R^{2}+Z^{2}}\nonumber\\
  &&+\big(\frac{2R^{2}-Z^{2}}{R^{2}+Z^{2}}
  +Z^{2}\big)\sin\sqrt{R^{2}+Z^{2}}\bigg].
\end{eqnarray}
Here $\gamma^{(0)}_{11}=\frac{e^{2}d^{2}_{x}\omega^{3}_0}{3\pi \hbar\varepsilon_0 c^{3}}$ is the spontaneous decay rate of atom in unbounded space and the dimensionless parameters $Z\equiv2z_0\omega_0/c$, $R\equiv r\omega_0/c$ are introduced for simplicity. The second part of $\gamma_{11}$ is the correction induced by the boundary and it can be seen that it is an oscillating function of the time required for a photon emitted by an atom to make a round trip between the atom and the boundary.
The first term of $\gamma_{12}$ as the vacuum term of modulation of the spontaneous emission rate is an oscillating function of the time needed by a photon to travel between the two atoms, the second term as the correction induced by the boundary is an oscillating function of the time required for a photon emitted by one atom to be reflected by the boundary and then be reabsorbed by another atom. In the evolution of entanglement the influence of one atom on another is reflected in $\gamma_{12}$.

 \begin{figure}[htbp]
\centering
\includegraphics[width=0.457\textwidth ]{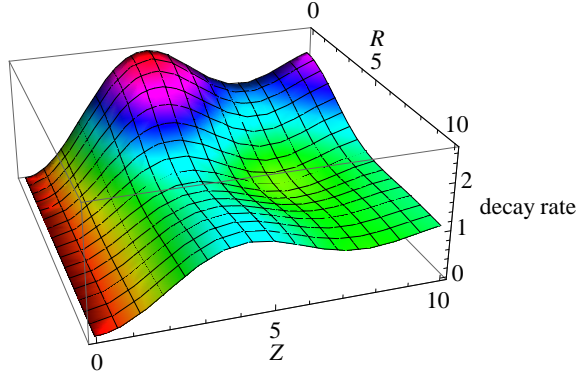}
\caption{(color online).  The decay rate of entanglement in the unit of $\gamma^{(0)}_{11}$ as a function of $R$ and $Z$ in the case of $x$ polarizations.
}\label{Fig.1}
\end{figure}
Although the concurrence between the two atoms decreases exponentially with time, as we show in (\ref{22}), but
when $Z$ is small, it can be found that the decay rate is proportional to $Z^{2}$.
So in the case that the boundary is placed very close to atoms, $Z\rightarrow0$, the decay rate will tend to zero, as is illustrated in Fig. $1$. Thus the entanglement can be totally protected and remain constant for a long time.
To show the efficiency of the presence of boundary, we find that when $Z$ takes $0.5, 0.1, 0.05$, respectively, the decay rate can reach to
 $5\times10^{-2}\gamma^{(0)}_{11}, 2\times10^{-3}\gamma^{(0)}_{11}$ and $5\times10^{-4}\gamma^{(0)}_{11}$.

\begin{figure}[htbp]
\centering
\includegraphics[width=0.457\textwidth ]{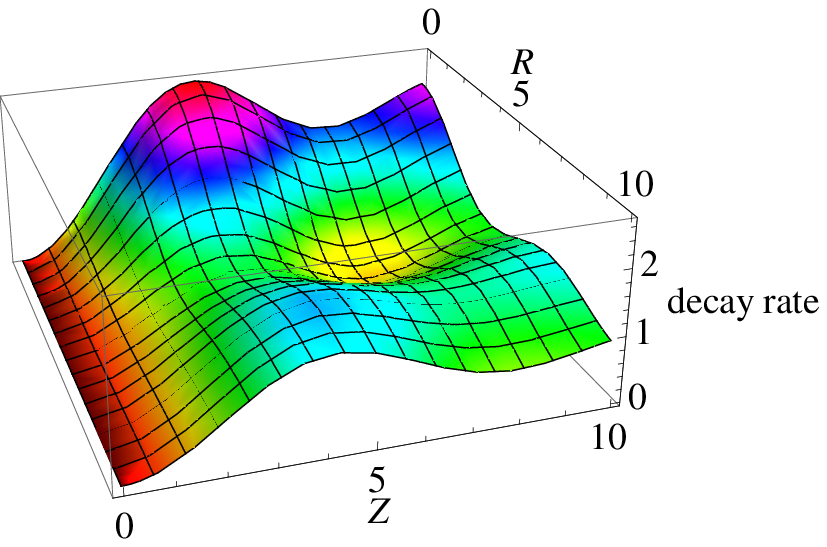}
\includegraphics[width=0.457\textwidth ]{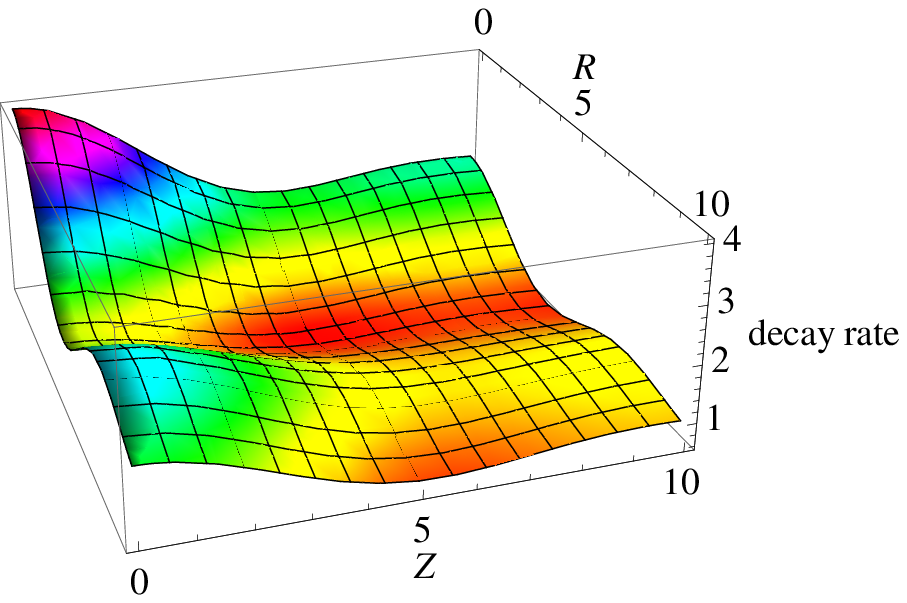}
\caption{(color online).  The decay rate of entanglement in the unit of $\gamma^{(0)}_{11}$ as a function of $R$ and $Z$ in the case of $y$ and $z$ polarizations respectively.
}\label{Fig.1}
\end{figure}
Previously, we only consider the polarizations of the two atoms in $x$-direction. Next we will apply the above developed formalism to other cases.
Here, let us note that the polarization directions of the atoms play a crucial role in the entanglement dynamics \cite{a} and
interatomic resonance interaction \cite{j}.
When the polarizations are along the $y$-axis, the spontaneous emission rate and the corresponding modulation can be written as
\begin{eqnarray}\label{f1}
\gamma_{11} &=&\gamma^{(0)}_{11}-3\gamma^{(0)}_{11}\frac{Z \cos Z+(Z^{2}-1)\sin Z}{2Z^{3}},\nonumber\\
  \gamma_{12} &=&3\gamma^{(0)}_{11}\frac{(R^{2}-1)\sin R+R \cos R}{2R^{3}}\nonumber\\&&-\frac{3\gamma^{(0)}_{11}}
  {2(R^{2}+Z^{2})}\bigg[\cos\sqrt{R^{2}+Z^{2}}+(\sqrt{R^{2}+Z^{2}}-\frac{1}{\sqrt{R^{2}+Z^{2}}})\sin\sqrt{R^{2}+Z^{2}}\bigg].
  \end{eqnarray}
In the case that $Z$ is small, the decay rate is also proportional
to $Z^{2}$. So the entanglement can also be shielded in the limit $Z\rightarrow0$.
For the polarizations in $z$-axis, similarly we have
\begin{eqnarray}\label{f2}
\gamma_{11} &=&\gamma^{(0)}_{11}-3\gamma^{(0)}_{11}\frac{Z \cos Z-\sin Z}{Z^{3}},\nonumber\\
  \gamma_{12} &=&3\gamma^{(0)}_{11}\frac{(R^{2}-1)\sin R+R \cos R}{2R^{3}}\nonumber\\&&+\frac{3\gamma^{(0)}_{11}}
  {2(R^{2}+Z^{2})^{3/2}}\bigg[\frac{R^{2}-2Z^{2}}{\sqrt{R^{2}+Z^{2}}} \cos\sqrt{R^{2}+Z^{2}}+\big(\frac{2Z^{2}-R^{2}}{R^{2}+Z^{2}}
  +R^{2}\big)\sin\sqrt{R^{2}+Z^{2}}\bigg].\nonumber\\
  \end{eqnarray}
It can be verified that in such a case $(\gamma_{11}+\gamma_{12})|_{Z\rightarrow0}=2((\gamma_{11}+\gamma_{12})|_{Z\rightarrow\infty}$.
This means that, when the atoms are placed near the boundary, the concurrence decays even faster than that without the boundary.
To show the role of polarizations clearly, we illustrate the decay rate of entanglement in $y$ and $z$ polarizations in Fig. $2$.

So only when the atoms are placed close to boundary and the polarizations of atoms is in the $xy$ plane can entanglement be protected for a long time. But why? This result can be attributed to
the fact that electric field should satisfy the boundary conditions on the surface of ideal plate: the tangential component of electric field on the surface should be zero. So when transversely polarizable atoms are close to the boundary, they can shielded from the influence of electric field vacuum fluctuations.

There is one point that deserves our attention: If we take $l_1$ norm \cite{norm} as the definition of coherence, it can be found that the entanglement in our case is just coherence. So our results also suit to the coherence for two atoms. In Ref. \cite{ta}, the coherence for two qubits system is investigated. Taking a close look, we can find that these two qubits have no interaction, so the modulation of spontaneous emission rate is not embedded in its expression. And if we put our two atoms very far from each other, then $\gamma_{12}=0$ and the decay rate will be $\gamma_{11}$ rather than its half as in \cite{ta}.

If the initial states of atoms is another Bell state $\psi^{-}=\frac{1}{\sqrt{2}}(|g_1e_2\rangle-|e_1g_2\rangle)$, following the same procedure, it can be found that the entanglement will decay with decay rate $\gamma_{11}-\gamma_{12}$. So when the two atoms are placed close to each other, the entanglement can also be protected, since in such a case $\gamma_{12}=\gamma_{11}$ as can be verified from their definitions.

 According to Dicke's theory \cite{dick, DeVoe}, the two two-level atoms can be treated as a single four-level system and the transition rate from collective states $\psi^{\pm}$ to ground state $|g_{1}g_{2}\rangle$ are superradiant decay rate $\Gamma_{\pm}$, respectively.
Meanwhile in this transition, the entanglement between atoms decrease from $1$ to $0$ with decay rate $\gamma_{11}\pm\gamma_{12}$. These two decay rates should be in direct proportion.
Besides, if we do not consider the influence of boundary and then compare $\gamma_{11}\pm\gamma_{12}$ in (\ref{f1}) or (\ref{f2}) with the formula (1) and FIG. 2. in \cite{DeVoe}, we can find  that $\gamma_{11}\pm\gamma_{12}$ are actually
 $\Gamma_{\pm}$, respectively. Thus we can identify $\gamma_{11}\pm\gamma_{12}$ with superradiant spontaneous emission rate. The appearance of
superradiance in our model stems from the fact that light emitted by one atom can be absorbed by
another atom, thus leading to cooperative processes in the emission \cite{dick, DeVoe, Eschne}.

In summary, we have investigated the dynamical evolution of entanglement between two polarizable two-level atoms in weak interaction with
a bath of fluctuating vacuum electromagnetic fields. Under the condition that the initial state
is Bell state $\psi^{\pm}$, we find that the entanglement between atoms decreases exponentially with decay rate equals to their collective superradiant decay rate, which depends on
 spontaneous emission rate of atom and the modulation of the spontaneous emission rate due to the presence of another identical atom.
It is shown that the entanglement is atomic polarizations and position dependent. When the polarizations of atoms are in the $xy$ plane and the distance between atoms and boundary is small, the entanglement will be protected for a very long time. It is all shown that the entanglement between atoms can also be protected if the atoms are in state $\psi^{-}$ and  put close enough.

We would like to thank L. Zhou for some help.

\end{document}